\documentclass[12pt,preprint]{aastex}
\makeatletter
  
  \@addtoreset{equation}{section}
\makeatother

\slugcomment{The Astrophysical Journal, Submitted 2002 August 3}

\shortauthors{Itoh et al.}
\shorttitle{ENHANCEMENT OF RESONANT THERMONUCLEAR REACTION RATES}

\received{2002 August 2}
\begin{document}

\title{ENHANCEMENT OF RESONANT THERMONUCLEAR REACTION RATES
 IN EXTREMELY DENSE STELLAR PLASMAS}

\author{\sc Naoki Itoh, Nami Tomizawa, and Shinya Wanajo}
\affil{Department of Physics, Sophia University,
       7-1 Kioi-cho, Chiyoda-ku, Tokyo, 102-8554, Japan;\\
       n\_itoh@sophia.ac.jp, tomiza-n@sophia.ac.jp, 
       wanajo@sophia.ac.jp}

\and

\author{\sc Satoshi Nozawa}
\affil{Josai Junior College for Women, 1-1 Keyakidai, Sakado-shi,
       Saitama, 350-0295, Japan; snozawa@josai.ac.jp}

\affil{The Astrophysical Journal, Submitted 2002 August 3}

\begin{abstract}

The enhancement factor of the resonant thermonuclear reaction rates is
calculated for the extremely dense stellar plasmas in the liquid phase.
In order to calculate the enhancement factor we use the screening
potential which is deduced from the numerical experiment of the classical
one-component plasma.  It is found that the enhancement is tremendous
for white dwarf densities if the $^{12}$C + $^{12}$C fusion cross sections
show resonant behavior in the astrophysical energy range.  We summarize
our numerical results by accurate analytic fitting formulae.

\end{abstract}



\section{INTRODUCTION}

In a recent important paper Cussons, Langanke, \& Liolios (2002) have
pointed out the potential resonant screening effects on stellar
$^{12}$C + $^{12}$C reaction rates.  The $^{12}$C + $^{12}$C fusion
cross sections show noticeable resonant structures down to the lowest
energies measured so far in the laboratory $E \sim 2.4$ MeV (Kettner,
Lorenz-Wirzba, \& Rolfs 1980).  If the resonant structure continues
to even lower energies and the astrophysical reaction rate is due to
the contributions of narrow resonances, one then has to consider that
the entrance channel width of these resonances will be modified in the
plasma.

Cussons, Langanke, \& Liolios (2002) have specifically pointed out the
possible importance of the plasma effects on the resonant $^{12}$C +
$^{12}$C reactions for a carbon white dwarf environment with $T = 5
\times 10^{7}$ K and $\rho = 2 \times 10^{9}$ g cm$^{-3}$.  They have
considered a resonance energy interval 0.4$-$2 MeV.  They have
specifically discussed a rather extreme case of the low resonance
energy $E_{r}$ = 400 keV and have estimated the overall enhancement
of the resonant $^{12}$C + $^{12}$C reaction rates due to the plasma
effects for this case.

Cussons, Langanke, \& Liolios (2002) adopted the method of Salpeter \&
Van Horn (1969) which is based on the lattice model of the dense plasma
to calculate the resonant screening effects.  One of the present authors
(N. I.) and his collaborators have calculated the enhancement of
non-resonant thermonuclear reaction rates in extremely dense stellar
plasmas (Itoh, Kuwashima, \& Munakata 1990).  This work is a natural
extension of the works of Itoh, Totsuji, \& Ichimaru (1977) and Itoh
et al. (1979), and improves upon the accuracy of the results of
Salpeter \& Van Horn (1969).  Itoh, Kuwashima, \& Munakata (1990) have
summarized their numerical results by an accurate analytical fitting
formula which will be readily implemented in the stellar evolution
computations.

The aim of the present paper is to extend the work of Itoh, Kuwashima,
\& Munakata (1990) to the case of resonant reactions.  The present paper
is organized as follows.  Physical conditions relevant to the present
calculation are made explicit in \S~2.  Calculation of the enhancement
factor of the resonant thermonuclear reaction rates is summarized in \S~3.
The results are presented in \S~4.  Extension to the case of ionic
mixtures is made in \S~5.  Concluding remarks are given in \S~6.

\section{PHYSICAL CONDITIONS}

First we consider thermonuclear reactions which take place in the plasma
in the thermodynamic equilibrium at temperature $T$ composed of one kind
of atomic nuclei and electrons with number densities $n_{i}$ and $n_{e}$
respectively; $Ze$ and $M$ denote the electric charge and the mass
of such an ion.  The conventional parameters which characterize such a
plasma are
\begin{eqnarray}
\Gamma & = & \frac{ \left(Ze \right)^{2} }{a k_{B} T} \, = \,
0.2275 \, \frac{Z^{2}}{T_{8}} \left( \frac{ \rho_{6}}{A} \right)^{1/3} \, , \\
\tau & = & \left[ \left( \frac{ 27 \pi^{2}}{4} \right) 
\frac{ M (Ze)^{4}}{ \hbar^{2} k_{B} T} \right]^{1/3} \, , 
\end{eqnarray}
where $a$ is the ion-sphere radius $a=(4 \pi n_{i}/3)^{-1/3}$, $A$ is the
mass number of the nucleus, $T_{8}$ is the temperature in units of
10$^{8}$ K, and $\rho_{6}$ is the mass density in units of 
10$^{6}$ g cm$^{-3}$.  In this paper we restrict ourselves to the case
that electrons are strongly degenerate.  This condition is expressed as
\begin{eqnarray}
T \ll T_{F} & = & 5.930 \times 10^{9} \, \left[ \left\{ 1 + 1.018 
 (Z/A)^{2/3} \rho_{6}^{2/3} \right\}^{1/2} - 1 \right] \,  {\rm K} \, ,
\end{eqnarray}
where $T_{F}$ is the electron Fermi temperature.  Furthermore, we consider
the case that the ions can be treated approximately as classical
particles.  The corresponding condition is written as
\begin{equation}
n_{i} \Lambda^{3} \leq 1 \, , 
\end{equation}
where $\Lambda = ( 2\pi \hbar^{2}/Mk_{B}T)^{1/2}$ is the thermal
de Broglie wave length of the ions.  The condition (2.4) is rewritten as
\begin{eqnarray}
T_{8} & \geq & 2.173 \rho_{9}^{2/3} / A^{5/3} \, ,
\end{eqnarray}
where $\rho_{9}$ is the mass density in units of 10$^{9}$ g cm$^{-3}$.
In this paper we impose a further condition that the ions are in the
liquid state (Slattery, Doolen, \& Dewitt 1982):
\begin{equation}
\Gamma < 178 \, .
\end{equation}

The parameter 3$\Gamma/\tau$ corresponds to the ratio of the classical
turning point radius at the Gamow peak and the mean interionic distance
in the case of the pure Coulomb potential (Itoh, Totsuji, \& Ichimaru
1977; Alastuey \& Jancovici 1978).  The theories of Itoh, Totsuji,
\& Ichimaru (1977), Itoh et al. (1979) and also Alastuey \& Jancovici
(1978) are valid under the condition 3$\Gamma/\tau \leq 1.6$.  Itoh,
Kuwashima, \& Munakata (1990) have extended the calculation of the
enhancement of the non-resonant thermonuclear reaction rates to the
case 3$\Gamma/\tau \leq 5.4$.  In Figure~1 we show the density-temperature
diagram of the pure $^{12}$C plasma which illustrates the physical
conditions described in this section.

\section{ENHANCEMENT OF THE RESONANT THERMONUCLEAR REACTION RATES}

In this paper we use the following screening potential for the
classical one-component plasma
\begin{eqnarray}
H(r) & = & C k_{B} T \, - \, \frac{Z^{2}e^{2}}{a} \left[
           \frac{1}{4} \left( \frac{r}{a} \right)^{2}
         - b \left( \frac{r}{a} \right)^{4}  \right] \, , \, 
\hspace{1.0cm} \, 0 \leq r \leq r_{0}  \\
H(r) & = & \frac{Z^{2}e^{2}}{a} \left[ 1.25 - \left( \frac{1.25}{2} \right)^{2} \, \frac{r}{a} \right] \, , \, \hspace{2.5cm} \,  r_{0} \leq r \leq 1.60 a  \\
C & = & 1.0531 \, \Gamma + 2.2931 \, \Gamma^{1/4} - 0.5551 \, \ln \Gamma
- 2.35 \, .
\end{eqnarray}
This screening potential is derived from the equilibrium pair correlation
function of the classical one-component plasma.  In adopting this
screening potential, our standpoint is the same as that of Alastuey \&
Jancovici (1978).  They have argued that the pair correlation function
should be taken as the static one.  The point is that the transmission
coefficient of the potential barrier is exceedingly small, which makes
nuclear reactions very rare events.  In a loose classical analogy, one
might say that, in most collisions, the colliding particles tunnel
through only a certain distance and are then reflected back.  Therefore,
as soon as $r$ is larger than a few nuclear diameters, equilibrium is
achieved and the probability of finding two nuclei at a distance $r$
from one another is given by the equilibrium pair correlation function.
Thus one can use the averaged potential in describing the tunnelling.
More recently DeWitt (1994), Rosenfeld (1994), and Isern \& Hernanz (1994)
have studied this problem.  They have essentially confirmed the
correctness of the method of Alastuey \& Jancovici (1978) on which our
work is based.  A similar work in the case of the solar fusion reactions
has been recently carried out by Bahcall et al. (2002) essentially
confirming the soundness of the method of the average potential.

  We also remark in relaion to the above-stated point that the nuclear
reactions as a whole are taking place on the macroscopic time scale,
whereas the screening potential is kept in equilibrium on the
microscopic time scale.

  We here remark on the validity of the classical one-component plasma
(OCP) model.  In the interior of dense stars the electron Fermi energy
becomes much larger than the Coulomb interaction energy between
the electron and the ion.  Therefore, the electron liquid becomes an
almost uniform liquid.  Owing to this fact, the interior of dense
stars can be satisfactorily described by the OCP model which consists
of point ions imbedded in the rigid background of electrons.
Of course, at the same density the OCP model becomes better for the
smaller ionic charge $Z$, as the ratio of the electron-ion interaction
energy to the electron Fermi energy becomes smaller for the smaller $Z$.

  The expression for $C$ is taken from Alastuey and Jancovici (1978).
The two segments of the screening potential are matched at $r=r_{0}$
so that the screening potential and its derivative be a continuous
function with respect to the distance $r$.  This procedure produces
solutions for $r_{0}$ and $b$ for the range of $\Gamma$-values
$4 \leq \Gamma \leq 90$.  Outside this range we use the value of $b$ which
makes (3.1) and (3.2) continuous at $r$=1.171875$a$.  In this case
the first derivatives of (3.1) and (3.2) are slightly discontinuous
at this point.  The linearly decreasing part
of the screening potential is identical to that employed by Itoh,
Totsuji, \& Ichimaru (1977) and also by Itoh et al. (1979).  The
screening potential (3.1) and (3.2) fits the results of the numerical
experiments excellently. (See Figure~1 of Itoh et al. (1979) for
the accuracy of this screening potential in reproducing the results
of the Monte Carlo computations.)  Note that this screening potential
exactly cancels the Coulomb potential $Z^{2}e^{2}/r$ at $r=1.60a$.
We further assume that the potential of mean force vanishes for
$r \geq 1.60a$.  Given the explicit form of the screening potential,
we are now in a position to calculate the enhancement of the
resonant thermonuclear reaction rates.

A single resonance in the cross section of a nuclear reaction
$0 + 1 \longrightarrow 2 + 3$ can be represented most simply as a
function of energy in terms of the classical Breit-Wigner formula
(Fowler, Caughlan, \& Zimmerman 1967)
\begin{eqnarray}
\sigma & = & \frac{ \pi \hbar^{2}}{2 \mu E} \, \frac{ \omega_{r} 
\Gamma_{1} \, \Gamma_{2}}{(E - E_{r})^{2} + 
  \frac{ \displaystyle{\Gamma_{tot}^{2}}}{ \displaystyle{4}}} \, ,
\end{eqnarray}
where $\mu$ is the reduced mass, $E$ is the center-of-mass energy,
$E_{r}$ is the resonance energy, $\omega_{r}$ is the statistical
weight factor, $\Gamma_{1}$ is the partial width for the decay of the
resonant state by reemission of $0 + 1$, $\Gamma_{2}$ is the partial
width for emission of $2 + 3$, $\Gamma_{tot}=\Gamma_{1}+\Gamma_{2}+\cdot
\cdot \cdot$ is the sum over all partial widths.  The partial width
$\Gamma_{1}$ is proportional to the barrier penetration factor $P(E)$
for the screened Coulomb potential.
\begin{eqnarray}
\Gamma_{1} & \propto & P(E) = \exp 
\left\{ -\frac{2 \sqrt{2 \mu}}{ \hbar}  \int_{0}^{r_{tp}} 
        \left[ V(r) - E \right]^{1/2} dr \right\} \, ,  \\
V(r) & = & \frac{Z^{2}e^{2}}{r} - H(r) \, ,
\end{eqnarray}
where $r_{tp}$ is the classical turning point radius which satisfies
the condition
\begin{equation}
V(r_{tp}) - E = 0 \, .
\end{equation}

We consider the case that the resonance is sharp; that is, the full width
at resonance, $\Gamma_{r}$, is considerably smaller than the effective
spread in energy of the interacting particles.  We further consider the
case $\Gamma_{1} \ll \Gamma_{2}$, $\Gamma_{tot} \approx \Gamma_{2}$.
Cussons, Langanke, \& Liolios (2002) have pointed out that for
$^{12}$C + $^{12}$C resonances far below the height of the Coulomb
barrier, the entrance channel width $\Gamma_{1}$ is much smaller than
the total resonance width.  The latter (which is of order $\sim$ 100 keV
for the observed resonances above 2.4 MeV) is also noticeably smaller
than the resonance energy.  In this case we have (Fowler, Caughlan,
\& Zimmerman 1967)
\begin{eqnarray}
\langle \sigma v \rangle & = & \left( \frac{2 \pi \hbar^{2}}{ \mu k_{B}T} \right)^{3/2}
\frac{ (\omega \gamma)_{r}}{ \hbar} \, \exp 
\left( - \frac{E_{r}}{k_{B}T} \right) \, ,  \\
 (\omega \gamma)_{r} & = & \omega_{r} \gamma_{r} = 
 \left( \frac{ \omega \Gamma_{1} \Gamma_{2}}{ \Gamma_{tot}} \right)_{r} \, 
\approx \, ( \omega \Gamma_{1} )_{r} \, .
\end{eqnarray}
Therefore the partial width $\Gamma_{1}$ in equation (3.5) is to be
evaluated at the resonance energy $E=E_{r}$.  Here we notice that the
resonance energy is shifted by the plasma effects.  We take $E_{r}$
to be the shifted resonance energy.  The shifted resonance energy $E_{r}$
is related to the resonance energy in the vacuum $E_{r}^{0}$ by the
relationship
\begin{eqnarray}
E_{r} &= & E_{r}^{0} \, - \, C \, k_{B} T  \, ,
\end{eqnarray}
where the expression for $C$ is given by equation (3.3).

The barrier penetration factor $P_{0}(E)$ for the pure Coulomb potential
$Z^{2}e^{2}/r$ of the identical nuclei is known to be
\begin{eqnarray}
P_{0}(E) & = & \exp \left\{ -2 \left[ \frac{a}
 {  \frac{ \displaystyle{ \hbar^{2}}}{ \displaystyle{MZ^{2}e^{2}} } }
 \right]^{1/2}
\frac{ \pi}{2} \frac{1}{ \sqrt{ \epsilon}}  \right\} \, ,  \\
\epsilon & = & \frac{a E}{Z^{2}e^{2}}  \, .
\end{eqnarray}
Therefore, the enhancement factor $\alpha$ of the resonant thermonuclear
reaction rates which arises because of the plasma screening effects is
\begin{eqnarray}
\alpha & = &  \exp \left\{ -2 \left[ \frac{a}
 {  \frac{ \displaystyle{ \hbar^{2}}}{ \displaystyle{MZ^{2}e^{2}} } } 
 \right]^{1/2} \left[ - \frac{ \pi}{2} \frac{1}{ \sqrt{ \epsilon_{r}^{0}}} +
 J(\Gamma, \epsilon_{r}) \right] \right\} \exp (C) \, ,  \\
 J(\Gamma, \epsilon) & = & \int_{0}^{x_{tp}} \left[ \frac{1}{x} -
  h(x) - \epsilon \right]^{1/2} dx \, ,  \\
 h(x) & = & \frac{a}{Z^{2}e^{2}} \, H(r) \, , \\
 x & = & \frac{r}{a} \, , \, \, \, \, x_{tp} \, = \, \frac{r_{tp}}{a} \, , \\
\epsilon_{r}^{0} & = & \frac{a E_{r}^{0}}{Z^{2}e^{2}}  \, ,  \, 
  \hspace{0.5cm} \, \epsilon_{r} \, = \, \frac{a E_{r}}{Z^{2}e^{2}} =
\epsilon_{r}^{0} - \frac{C}{ \Gamma} \, .
\end{eqnarray}
Here we notice that our method is valid for $\epsilon_{r}=\epsilon_{r}^{0}
- \frac{C}{ \Gamma} \geq 0.$

Since we have (Itoh et al. 2002)
\begin{eqnarray}
a & = & 0.7346 \times 10^{-10} \left( \frac{ \rho_{6}}{A} 
            \right)^{-1/3} \, \, {\rm cm} \, ,
\end{eqnarray}
we can rewrite equation (3.13) as
\begin{eqnarray}
\alpha & = &  \exp \left\{ -1.004 \times 10^{1} Z A^{2/3} 
  \rho_{6}^{-1/6} \, \left[ - \frac{ \pi}{2} \frac{1}{ \sqrt{ 
  \epsilon_{r} + \frac{C}{ \Gamma}}} + J(\Gamma, \epsilon_{r}) 
  \right]  \right\} \exp (C) \, \nonumber  \\
  & \equiv &  \exp \left[ -1.004 \times 10^{1} Z A^{2/3} 
  \rho_{6}^{-1/6} \, K(\Gamma, \epsilon_{r}) \right] \exp (C) \, .
\end{eqnarray}
We also have a useful relationship
\begin{eqnarray}
\frac{Z^{2}e^{2}}{a} & = & 1.960 \times 10^{-3} \, Z^{2} 
\left( \frac{ \rho_{6}}{A} \right)^{1/3} \, {\rm MeV} \, .
\end{eqnarray}
This gives $Z^{2}e^{2}/a=0.308$ MeV and
1.004$\times$10$^{1}Z A^{2/3} \rho_{6}^{-1/6}$ = 99.85 for $Z=6$, $A=12$,
$\rho=10^{9} \, {\rm g \, cm}^{-3}$.

\section{RESULTS}

We have carried out the numerical integration of $J(\Gamma, \epsilon)$
in equation (3.14) for various values of $\Gamma$ and $\epsilon$.
In Figure~2 we show the function $J(\Gamma, \epsilon)$ as a function
of $\epsilon$ for various values of $\Gamma$.  In Figure~3 we show
the function $K(\Gamma, \epsilon_{r})$ as a function of $\epsilon_{r}$
for various values of $\Gamma$.

In order to facilitate the application of the numerical results
obtained in the present paper we will present an accurate analytic
fitting formula for $K(\Gamma, \epsilon_{r})$.  We have carried out
the numerical calculations of $K(\Gamma, \epsilon_{r})$ for
$1 \leq \Gamma \leq 200$, $0 \leq \epsilon_{r} \leq 10$.  We express
the analytic fitting formula by
\begin{eqnarray}
\log_{10} K(\Gamma, \epsilon_{r}) & = & \sum_{i,j=0}^{10} a_{i j} \, g^{i} u^{j} \, ,  \\
 g & \equiv & \frac{1}{1.1505} \left(\log_{10} \Gamma \, - \,
              1.1505 \right) \, ,  \\
 u & \equiv & \frac{1}{5.0} \left( \epsilon_{r} \, - \, 5.0 \right) \, .
\end{eqnarray}
The coefficients $a_{ij}$ are presented in Table~1.  The accuracy of the
fitting is generally better than 0.1\%.

\section{IONIC MIXTURES}

Itoh et al. (1979) analyzed the results of the Monte Carlo computations
for the screening potentials of the ionic mixtures of various charge
ratios and concentration ratios carried out by them and also by Hansen,
Torrie, \& Vieillefosse (1977).  They have established that the
screening potential at intermediate distances $H_{ij}(r)$ for a mixture
of two kinds of ions with charges and number densities $(Z_{1}e, n_{1})$
and $(Z_{2}e, n_{2})$ given below fits the Monte Carlo results excellently
within the inherent Monte Carlo noise:
\begin{eqnarray}
\frac{ H_{i j}(r)}{k_{B}T \Gamma_{ij}} & = & 1.25 \, - \, 0.390625 \,
  \frac{r}{(a_{i}+a_{j})/2}  \, \hspace{1.0cm} \, (i, j= 1, 2) \, , \\
\Gamma_{ij} & = & \frac{ Z_{i} Z_{j}e^{2} }{(1/2)(a_{i}+a_{j})k_{B}T}
   \, \hspace{2.8cm} \, (i, j= 1, 2) \, ,  \\
a_{1} & = & \left[ \frac{3 Z_{1}}{4 \pi (Z_{1}n_{1} + Z_{2}n_{2})} 
            \right]^{1/3} \, , \\
a_{2} & = & \left[ \frac{3 Z_{2}}{4 \pi (Z_{1}n_{1} + Z_{2}n_{2})} 
            \right]^{1/3} \, .
\end{eqnarray}
We also define the parameter
\begin{eqnarray}
\tau_{ij} & = & \left[ \left( \frac{27 \pi^{2}}{4} \right)
  \frac{2 \mu_{ij} Z_{i}^{2} Z_{j}^{2} e^{4}}{ \hbar^{2} k_{B}T}
  \right]^{1/3} \, \hspace{2.5cm}  \, (i, j= 1, 2) \, ,
\end{eqnarray}
where $\mu_{ij}$ is the reduced mass for the two ions with charges
$Z_{i}$ and $Z_{j}$.  Then the enhancement factor for the resonant
thermonuclear rates of the two nuclei $Z_{i}$ and $Z_{j}$ is given by
\begin{eqnarray}
\alpha & = &  \exp \left\{ -2 \left[ \frac{ \displaystyle{
  \frac{1}{2} (a_{i}+a_{j})}}
 {  \frac{ \displaystyle{ \hbar^{2}}}{ 
    \displaystyle{2 \mu_{ij} Z_{i} Z_{j} e^{2}} } } \right]^{1/2} 
 \left[- \frac{ \pi}{2} \frac{1}{ \sqrt{ \epsilon_{r} 
           + \frac{C_{ij}}{ \Gamma_{ij}}}} +
 J(\Gamma_{ij}, \epsilon_{r}) \right] 
    \right\} \exp (C_{ij}) \,    \nonumber \\
& \equiv &  \exp \left\{ -2 \left[ \frac{ \displaystyle{ 
  \frac{1}{2} (a_{i}+a_{j}) }}
 {  \frac{ \displaystyle{ \hbar^{2}}}{ 
    \displaystyle{2 \mu_{ij} Z_{i} Z_{j} e^{2}} } } \right]^{1/2} 
 K(\Gamma_{ij}, \epsilon_{r}) \right\} \exp (C_{ij}) \, ,  \\
\epsilon_{r} & = & \frac{1}{2} \frac{ (a_{i}+a_{j}) E_{r}}
{ Z_{i} Z_{j} e^{2}}  \, = \, \frac{1}{2} \frac{ (a_{i}+a_{j}) E_{r}^{0}}
{ Z_{i} Z_{j} e^{2}} \, - \, \frac{C_{ij}}{ \Gamma_{ij}}  \, ,  \\
 C_{ij} & = & 1.0531 \, \Gamma_{ij} + 2.2931 \, \Gamma_{ij}^{1/4}
  - 0.5551 \, \ln \Gamma_{ij} - 2.35 \, .
\end{eqnarray}
The analytic fitting formula for this case $K(\Gamma_{ij}, \epsilon_{r})$
has the same form as equations (4.1), (4.2), (4.3).

\section{CONCLUDING REMARKS}

We have presented a calculation of the enhancement of the resonant
thermonuclear reaction rates for extremely dense stellar plasmas.
The calculation has been carried out by adopting the screening potential
derived from the Monte Carlo computations of the classical
one-component plasma.  We have summarized our numerical results
by an accurate analytic fitting formula to facilitate applications.
The present results will be useful if the $^{12}$C + $^{12}$C fusion
reaction contains narrow resonances in the astrophysical energy range.

\acknowledgments

We wish to thank K. Langanke for making the preprint available to us
prior to its publication.  We also thank Y. Oyanagi for allowing us
 to use the least-squares fitting program SALS.  We are most grateful
to our referee for many valuable comments on the original manuscript
which helped us tremendously in revising the manuscript.  This work
is financially supported in part by Grants-in-Aid of the Japanese
Ministry of Education, Culture, Sports, Science, and Technology
under contracts 13640245, 13740129.

\clearpage

\begin{table*}
\footnotesize
\caption[]{Coefficients $a_{ij}$}
\begin{tabular}{crrrrrr} \hline

  & {\it j}=0 \, \, \, & {\it j}=1 \, \, \, & {\it j}=2  \, \, \, & 
    {\it j}=3  \, \, \, & {\it j}=4  \, \, \, & {\it j}=5  \, \, 
    \, \\ \hline

 {\it i}=0 &     $-$3.65927E+0 & $-$1.26734E+0 &  6.07584E$-$1 &  6.30602E$-$2 &
 $-$7.78163E$-$1 & $-$1.79849E+0 \\

 {\it i}=1 &      1.55574E$-$3 & $-$4.55740E$-$3 &  2.58322E$-$2 &  7.00779E$-$2 &
 $-$2.64033E$-$1 & $-$3.01686E$-$1 \\

 {\it i}=2 &      6.20281E$-$3 & $-$1.79841E$-$2 &  1.04425E$-$1 &  2.79543E$-$1 &
 $-$1.07405E+0 & $-$1.21880E+0 \\

 {\it i}=3 &     $-$7.53523E$-$4 &  3.65732E$-$2 & $-$2.62102E$-$1 & $-$7.31969E$-$1 &
 2.74892E+0 &  3.19750E+0 \\

 {\it i}=4 &      1.18877E$-$2 &  1.11979E$-$2 & $-$1.05106E$-$1 & $-$4.62001E$-$1 &
 1.32482E+0 &  2.26167E+0 \\

 {\it i}=5 &     $-$4.10209E$-$2 & $-$9.73313E$-$2 &  8.96704E$-$1 &  2.55956E+0 &
 $-$9.59895E+0 & $-$1.13011E+1 \\

 {\it i}=6 &      1.89159E$-$2 &  5.88188E$-$2 & $-$5.47015E$-$1 & $-$1.22948E+0 &
 5.38904E+0 &  4.80222E+0 \\

 {\it i}=7 &      5.10796E$-$2 &  1.10523E$-$1 & $-$1.09263E+0 & $-$2.99190E+0 &
 1.16478E+1 &  1.32623E+1 \\

 {\it i}=8 &     $-$4.26759E$-$2 & $-$1.08244E$-$1 &  1.02972E+0 &  2.53992E+0 &
 $-$1.04973E+1 & $-$1.05685E+1 \\

 {\it i}=9 &     $-$1.97340E$-$2 & $-$4.38567E$-$2 &  4.51951E$-$1 &  1.19137E+0 &
 $-$4.79595E+0 & $-$5.30281E+0 \\

 {\it i}=10 &      2.00491E$-$2 &  5.10687E$-$2 & $-$4.93422E$-$1 & $-$1.22559E+0
 &  5.06476E+0 &  5.18690E+0 \\ \hline

\end{tabular}

\smallskip

\begin{tabular}{crrrrr} \hline

  & {\it j}=6 \, \, \, & {\it j}=7 \, \, \, & {\it j}=8  \, \, \, & 
    {\it j}=9  \, \, \, & {\it j}=10   \, \, \, \\ \hline

 {\it i}=0 &     3.53738E+0 &  2.90085E+0 & $-$4.98720E+0 & $-$1.80622E+0 &
 2.62678E+0 \\

 {\it i}=1 &     8.84245E$-$1 &  4.72177E$-$1 & $-$1.18370E+0 & $-$2.24453E$-$1 &
 5.27141E$-$1 \\

 {\it i}=2 &     3.62099E+0 &  1.92629E+0 & $-$4.87413E+0 & $-$9.25366E$-$1 &
 2.18368E+0 \\

 {\it i}=3 &    $-$9.25572E+0 & $-$5.12232E+0 &  1.25061E+1 &  2.53278E+0 &
 $-$5.66738E+0 \\

 {\it i}=4 &    $-$4.95541E+0 & $-$4.03418E+0 &  7.40638E+0 &  2.26404E+0 &
 $-$3.72743E+0 \\

 {\it i}=5 &     3.24488E+1 &  1.84055E+1 & $-$4.41392E+1 & $-$9.35145E+0 &
 2.02577E+1 \\

 {\it i}=6 &    $-$1.70172E+1 & $-$6.83208E+0 &  2.14258E+1 &  2.91262E+0 &
 $-$8.99871E+0 \\

 {\it i}=7 &    $-$3.92927E+1 & $-$2.16796E+1 &  5.33707E+1 &  1.10456E+1 &
 $-$2.44789E+1 \\

 {\it i}=8 &     3.41813E+1 &  1.61786E+1 & $-$4.46244E+1 & $-$7.64260E+0 &
 1.95906E+1 \\

 {\it i}=9 &     1.61526E+1 &  8.69953E+0 & $-$2.19164E+1 & $-$4.44246E+0 &
 1.00465E+1 \\

 {\it i}=10 &    $-$1.66155E+1 & $-$8.08293E+0 &  2.18847E+1 &  3.90128E+0 &
 $-$9.71024E+0 \\ \hline

\end{tabular}
\end{table*}

\clearpage

\begin{figure}
\epsscale{0.8}
\plotone{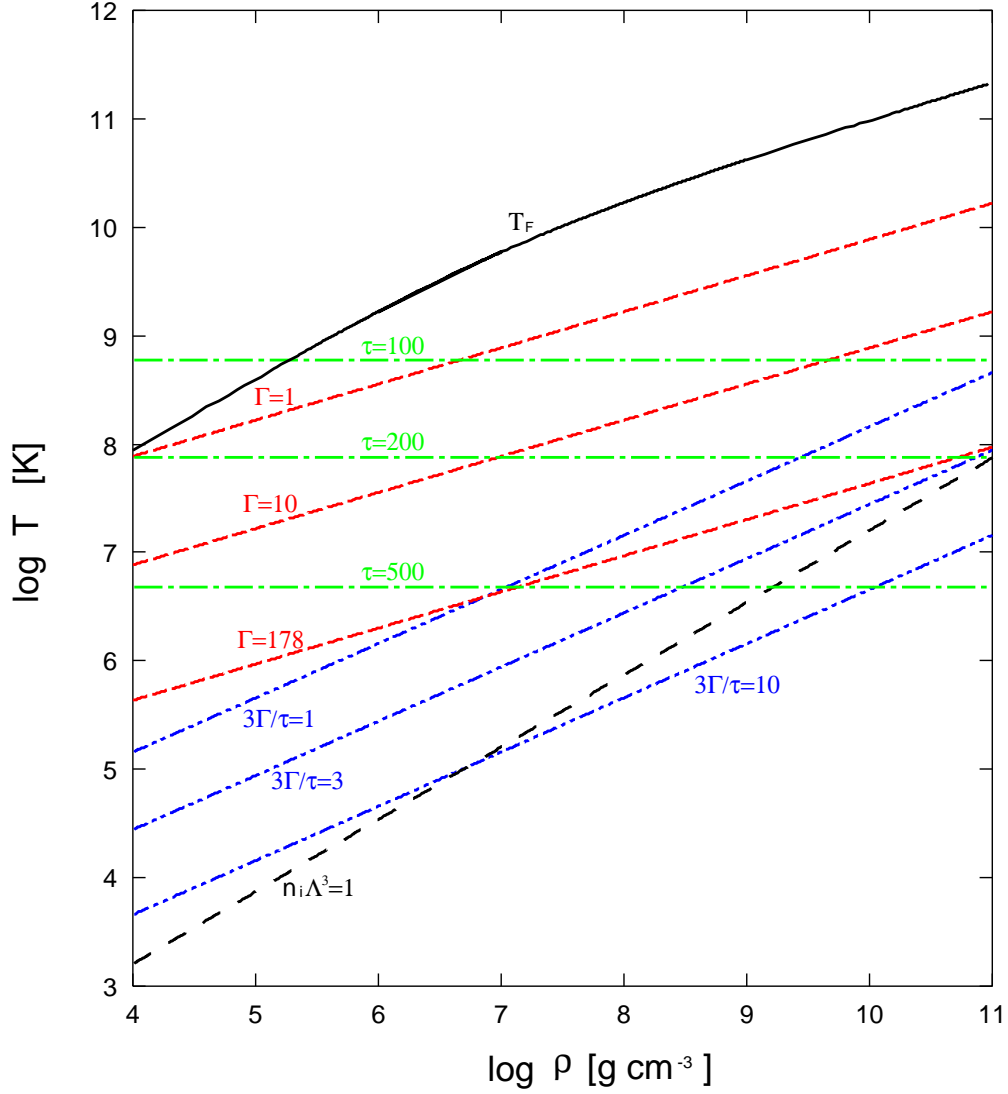}
\caption{Physical conditions for the pure $^{12}$C plasma.  The present
calculation is valid in the region bounded by the electron Fermi
temperature $T_{F}$ and the line $n_{i} \Lambda^{3}=1$.  Furthermore,
the ions should be in the liquid state: $\Gamma < 178$.  The lines of
$\tau$=const., $\Gamma$=const., and $3 \Gamma/\tau$=const. are shown.}
\end{figure}

\clearpage
\begin{figure}
\epsscale{0.8}
\plotone{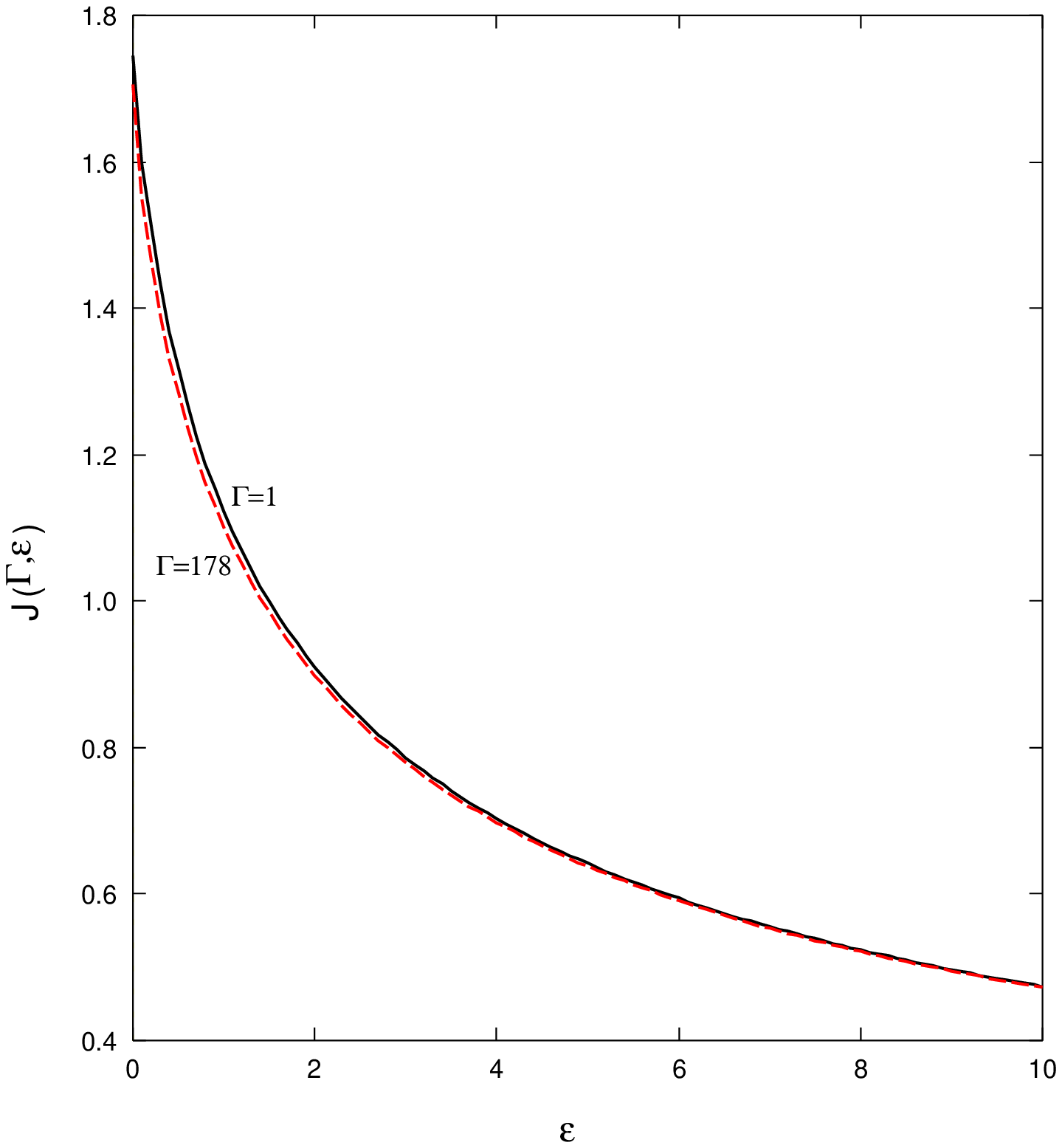}
\caption{$J(\Gamma, \epsilon)$ as a function of $\epsilon$ for
various values of $\Gamma$.}
\end{figure}

\clearpage
\begin{figure}
\epsscale{0.8}
\plotone{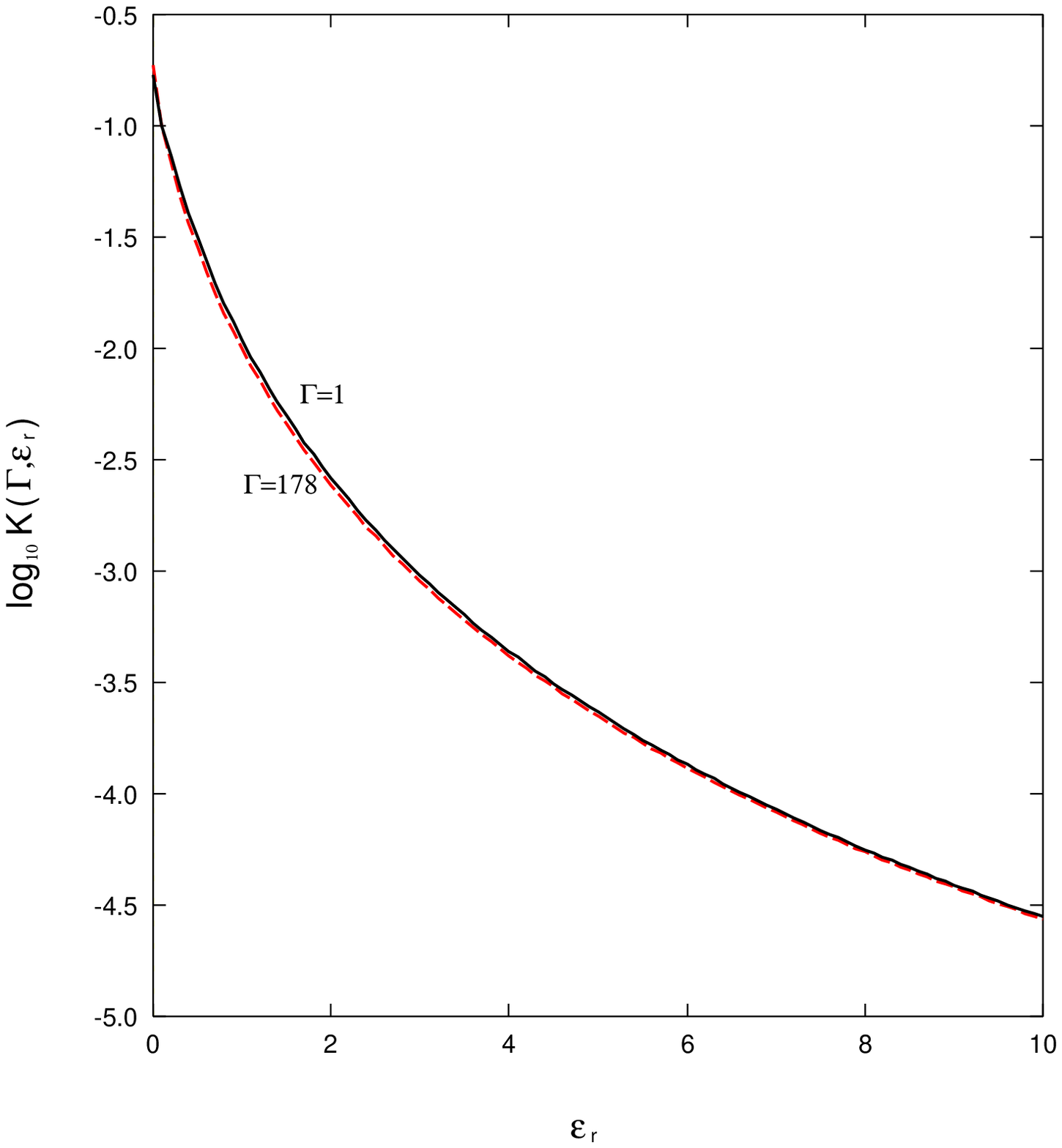}
\caption{$K(\Gamma, \epsilon_{r})$ as a function of $\epsilon_{r}$
for various values of $\Gamma$.}
\end{figure}

\end{document}